\documentclass[11pt]{article}

\usepackage{stolenstyle}

\usepackage{amsmath,epsf,amssymb,latexsym,amsthm,setspace,bbm,array,pifont}

\usepackage[matrix,arrow,frame,import,curve,color]{xy}

\DeclareFontFamily{U}{rsf}{}
\DeclareFontShape{U}{rsf}{m}{n}{
  <5> <6> rsfs5 <7> <8> <9> rsfs7 <10-> rsfs10}{}
\DeclareMathAlphabet\Scr{U}{rsf}{m}{n}

\def\CO#1#2{{[#1,#2]}}
\def\AC#1#2{{\{#1,#2\}}}

\def\cDb{{\overline{\cD}}}
\def\cQb{{\overline{\cQ}}}

\def\P{{\mathbb P}}

\def\Ker{\operatorname{Ker}}

\def\Tr{\operatorname{Tr}}

\def\ch{\operatorname{ch}}

\def\GU{\operatorname{U{}}}

\def\p{\partial}

\def\la{\langle}
\def\ra{\rangle}

\def\ff#1#2{{\textstyle\frac{#1}{#2}}}
\def\half{\frac{1}{2}}

\def\cC{{\cal C}}
\def\cD{{\cal D}}
\def\cE{{\cal E}}

\def\cK{{\cal K}}
\def\cL{{\cal L}}

\def\cO{{\cal O}}

\def\cQ{{\cal Q}}
\def\cR{{\cal R}}

\def\cV{{\cal V}}

\newcommand\gammab{\overline{\gamma}}

\newcommand\etab{\overline{\eta}}
\newcommand\thetab{\overline{\theta}}

\newcommand\lambdab{\overline{\lambda}}

\newcommand\xib{\overline{\xi}}

\newcommand\sigmab{\overline{\sigma}}
\newcommand\taub{\overline{\tau}}

\newcommand\phib{\overline{\phi}}
\newcommand\chib{\overline{\chi}}
\newcommand\psib{\overline{\psi}}

\newcommand\Gammab{\overline{\Gamma}}

\newcommand\Lambdab{\overline{\Lambda}}
\newcommand\Xib{\overline{\Xi}}

\newcommand\Sigmab{\overline{\Sigma}}
\newcommand\Upsilonb{\overline{\Upsilon}}
\newcommand\Phib{\overline{\Phi}}

\newcommand\Eb{\overline{E}}
\newcommand\Fb{\overline{F}}
\newcommand\Gb{\overline{G}}
\newcommand\Hb{\overline{H}}

\newcommand\Jb{\overline{J}}
\newcommand\Kb{\overline{K}}

\newcommand\Pb{\overline{P}}

\newcommand\Sb{\overline{S}}

\def\pbb{{\overline{p}}}
\def\sb{{\overline{s}}}

\def\cOb{{\overline{\cO}}}

\def\preprint{{{NSF-KITP-14-158}}}

\title{Worldsheet instantons and (0,2) linear models}
\author[a,b]{Marco Bertolini}
\author[a] {and M.~Ronen Plesser}
\affiliation[a]{Center for Geometry and Theoretical Physics, Box 90318 \\
Duke University, Durham, NC 27708-0318, USA}
\affiliation[b]{Kavli Institute for Theoretical Physics\\
University of California, Santa Barbara, CA 93106-4030, USA}
\emailAdd{mb266@phy.duke.edu}
\emailAdd{plesser@cgtp.duke.edu}
\abstract{We study the stability of heterotic compactifications described
  by (0,2) gauged linear sigma models with respect to worldsheet instanton
corrections to the space-time superpotential following the work of
Beasley and Witten \cite{Beasley:2003fx}.
We show that generic models elude the vanishing theorem proved there,
and may not determine supersymmetric heterotic vacua.
We then construct a subclass of linear models for which a vanishing
theorem holds, generating an extensive list of consistent 
heterotic backgrounds.
}

\begin{document}

\maketitle

\section{Introduction}\label{s:intro}

A natural starting point for exploring the moduli space of (0,2)
heterotic compactifications is the study of the geometry of
holomorphic vector bundles $\cV$ over Calabi-Yau (CY) manifolds $M$.
Under suitable conditions such bundles determine, to all orders in
$\alpha'$, a supersymmetric heterotic vacuum.  For a long time it has
been known \cite{Dine:1986zy} that worldsheet instantons wrapping
rational curves in $M$ in principle generate a potential which
destabilizes the vacuum.  
In rather special cases, such as models with (2,2) supersymmetry
\cite{Dixon:1987bg} or some specially
 fine-tuned (0,2) models \cite{Distler:1986wm,Distler:1987ee}, the
 correction terms vanish for each instanton separately, but this is
 not true in more generic models \cite{Berglund:1995yu,Braun:2007xh,Aspinwall:2011us}.

 In this context, heterotic compactifications obtained as gauged
 linear sigma models (GLSMs) \cite{Witten:1993yc} have received
 special attention, as they are believed to be stable under
 worldsheet instantons, even in the generic case in which the
 contributions of individual instantons do not vanish.  This claim is
 then a nontrivial vanishing theorem about the total contribution from
 each instanton class.   This was first proposed in
 \cite{Silverstein:1995re} and further studied in
 \cite{Beasley:2003fx}.\footnote{In \cite{Basu:2003bq} an argument
   along rather different lines was pursued.}  This work
 suggests that in these models the corrections vanish even when the
 contributions of individual instantons do not.  If true, this would
 guarantee the existence of a vast playground for tackling issues of
 (0,2) moduli spaces.  

 Essentially, these arguments rely on the fact that in heterotic vacua
 determined by (0,2) theories the space-time superpotential for gauge singlets can be
 determined by a correlator ${\cal C}$ computed in a (half-) twisted
 version of the model.  Unlike the twisted versions of (2,2) theories,
 this is not a topological field theory, but the simplifications
 associated to the existence of a nilpotent scalar charge, such as the
 decoupling of exact operators from the correlators of closed
 operators, carry over to this case and show that ${\cal C}$ depends
 holomorphically on the relevant worldsheet couplings.

 This holomorphy together with compactness arguments can be used
 to show that the correlator vanishes identically.  The argument of
 \cite{Silverstein:1995re} used the fact that the parameter space of
 the GLSM is compact (or has a natural compactification).  ${\cal C}$
 was shown to be a global section of a holomorphic bundle of negative
 curvature.  If nonzero this must exhibit poles, which in this theory
 arise from the finite-energy configurations with very large field
 values which occur at special loci in the parameter space.  At these
 loci the model is indeed singular, but the large-field region can be
 studied semiclassically to demonstrate that these configurations do
 not lead to any singularities in ${\cal C}$.  The absence of poles
 shows that this vanishes identically.  This was pursued explicitly in
 a simple example, but the argument did not appear to rely on details
 of this example so seemed likely to generalize.

 In turn, the argument of \cite{Beasley:2003fx}
 relied on the compactness of an appropriate moduli space of
 instantons.  More precisely, these authors used the fact that
 the contribution to ${\cal C}$ at any fixed instanton number can be
 related to a calculation in a model in which many of the worldsheet
 couplings vanish.  In this model, the moduli space of instantons is
 compact, and a zero-mode counting argument shows that the
 contribution to ${\cal C}$ vanishes.   Here too, detailed calculations
 were done in simple examples but the argument seemed very robust and
 likely to hold in general.

In this note, we follow up on this work with a systematic study of the
conditions under which the vanishing theorem of \cite{Beasley:2003fx}
applies.  We find that in a generic gauged linear sigma model the
argument that the moduli space of instantons is compact
fails, reviving the question of whether these models are in
fact destabilized by worldsheet instantons.  We do not resolve this
question.  We are, however, able to construct an extensive class of
models for which the argument holds -- a sizable playground, if not as
extensive as had been hoped.

The rest of the paper is organized as follows. In Section \ref{s:model} we review the construction of (0,2) linear models relevant for our analysis.
In Section \ref{s:examples} we show that there exist models for which the vanishing 
of the space-time superpotential for gauge singlets is not guaranteed and we present an example in detail.
In Section \ref{s:theorem} we prove a vanishing theorem for a particular subclass of (0,2) linear models.
In Section \ref{s:conclusions} we end with some implications of this work and future directions.

\acknowledgments We would like to thank Nick Addington, Paul
Aspinwall, Chris Beasley, Ilarion Melnikov, Eric Sharpe and Edward Witten for useful conversations.
MB would like to particularly thank Ilarion Melnikov for pointing out
the problem and for countless discussions.  This work is supported in
part by the NSF Grant PHY-1217109 (MB and MRP) and NSF Grant
PHY11-25915 (MB).  We thank the organizers of the workshop on
Heterotic Strings and (0,2) QFT at Texas A\&M University in May, 2014
where this project began.  MRP thanks the particle physics group at Tel
Aviv University, and especially Yaron Oz, for their gracious hospitality
while pursuing this work.

\section{The linear model}
\label{s:model}

Our tool for investigating the issue of instanton corrections in this
note is the (0,2) gauged linear sigma model.  For a suitably
constructed bundle $\cV$ on a CY space $M$ presented as a complete
intersection $H_A = 0$ in a Fano toric variety $V$,\footnote{We recall
  that a variety $V$ is Fano if and only if the anticanonical bundle
  $K_V$ of $V$ is ample.} the IR worldsheet dynamics is expected to be
the same as that of an Abelian gauge theory with (0,2) supersymmetry.
In this section we are going to review the construction of the (0,2)
linear model \cite{Witten:1993yc} in order to establish notation and
define the class of models we consider. More details can be found in
appendix \ref{app:linearmodel}.

The linear models we consider are gauge theories with gauge group
$\GU(1)^R$, along with $m$ neutral chiral supermultiplets we call
$\Sigma_\mu=(\sigma_\mu,\lambda_{\mu,+})$.  We couple these to 
a collection of charged supermultiplets determined by the
geometric data:
\begin{align}
\xymatrix@R=1mm@C=3mm{
\text{fields}			&P^\alpha			&\Phi^i		&\Gamma^I		&\Lambda^A		&S			&\Xi			\\
\GU(1)^a				&-m_\alpha^a		&q_i^a		&Q^a_I			&-d^a_A			& m^a-d^a	& d^a-m^a 		\\
}
\end{align}
where 
\begin{align}
m^a& = \sum_\alpha m_\alpha^a~, &d^a &= \sum_A d_A^a~.
\end{align} 

The $n$ chiral multiplets $\Phi^i=(\phi^i,\psi^i)$ and their charges are determined by a
presentation of $V$ as a symplectic $\GU(1)^R$ quotient.   The model has
Fayet-Iliopoulos D-terms whose values $r^a$ correspond to the shift in the
moment map for the $\GU(1)^R$ action.  The moduli space of classical
vacua of a theory containing only these fields will be $V$ when the
$r^a$ lie in a cone $\cK_V$, the K\" ahler cone of $V$.

The $N$ Fermi multiplets $\Gamma^I$, with lowest components the left-moving fermions $\gamma^I$, satisfy a chirality condition 
 \footnote{Unless otherwise specified, we use Einstein's summation convention throughout the paper.}
\begin{align}
\label{eq:Ecoup1}
\cDb \Gamma^I &= \sqrt2 E^I(\Sigma,\Phi)~, 
&E^{I}(\Sigma,\Phi) &=\Sigma_\mu E^{I\mu}(\Phi)~,
\end{align}
and their charges determine the bundle $\cE\rightarrow V$ by the short exact sequence (SES)
\begin{align}
\label{eq:SESbundleV}
\xymatrix@R=0mm@C=10mm{
0  \ar[r]&\oplus_\mu \cO \ar[r]^-{E^{I\mu}}  &\oplus_I \cO(Q_I) \ar[r]  & \cE   \ar[r] &0~.
}
\end{align}
This collection of fields with these couplings comprises what we refer
to as the $V$ model \cite{Morrison:1994fr}.  It is not a conformal field theory, and will
typically exhibit trivial IR behavior.  The space of $V$ models is
parameterized by $r^a$ complexified by $\theta$ angles as well
as the coefficients of the maps $E^{I\mu}$.\footnote{These are in general
subject to identifications, so this is an overparameterization.}

In general, there is a larger cone, which we call the {\sl geometric
  cone\/}  $\cK_c$, in which the space of vacua has this character.
More precisely, this is the cone in which the $V$ model as defined
above has supersymmetric classical vacua.  It is generated by $q_i$,
and it is divided into phases. These correspond to subcones of
$\cK_c$, one of which is $\cK_V$, separated by
hyperplanes associated to $\GU(1)$ subgroups of the gauge group 
which are unbroken at large $\phi$.

To construct our superconformal theory we augment the $V$ model by
the $k$ chiral multiplets $P^\alpha=(p^\alpha,\chi^\alpha)$.  For $r\in\cK_V$ the
space of classical vacua for the theory including these is the total
space $V^+ = \text{tot}\left( \oplus_\alpha\cO(-m_\alpha)\rightarrow V\right)$.
We introduce as well $L$ Fermi multiplets $\Lambda^A$, whose lowest components are the left-moving fermions $\eta^A$, satisfying a
chirality condition 
\begin{align}
\label{eq:Ecoup2}
\cDb \Lambda^A &= \sqrt2 E^A(P,\Sigma,\Phi)~, 
&E^A(P,\Sigma,\Phi)&= \Sigma_\mu P^\alpha E^{A\mu}_\alpha(\Phi)~.
\end{align}
The model with these fields and couplings will be referred to as the
$V^+$ model.  Like the $V$ model, it will not in general be
conformal.  In addition to the parameters listed above, it is
specified by the coefficients of the maps $E^{A\mu}_\alpha$.

The conformal model in which we are interested -- the M model -- is
obtained from the $V^+$ model by adding a superpotential interaction
\begin{align}
\int d\theta^+ \left( \Lambda^A H_A(\Phi) + \Gamma^I J_{I}(P,\Phi)\right)\big|_{\thetab^+=0}  +\text{h.c.} ~,
\end{align}
where
\begin{align}
\label{eq:Jsup}
J_I(P,\Phi)=P^\alpha J_{I\alpha}(\Phi)~,
\end{align}
subject to the conditions
\begin{align}
\label{eq:SUSYcond}
\sum_A H_A E^{A\mu}_\alpha +\sum_I J_{I\alpha} E^{\mu I}&=0\qquad \forall \alpha,\mu~,  
\end{align}
required in order to preserve (0,2) supersymmetry.  For $r\in\cK_V$ and generic $H_A$,
the space of classical vacua is the complete intersection $M =
\{\phi\in V|H_A(\phi)=0\}$.  When this is nonsingular the $\Lambda$
fermions all acquire a mass and the light left-moving fermions take
values in the bundle $\cV\rightarrow M$ defined by the restriction to $M$ of
the complex
\begin{align}\label{eq:cvseq}
\xymatrix@R=0mm@C=10mm{
0 \ar[r]  &\oplus_\mu\cO \ar[r]^-{E^{I\mu}} & \oplus_I \cO(Q_I) \ar[r]^-{J_{I\alpha}} & \oplus_\alpha \cO(m_\alpha) \ar[r] &0~,
}
\end{align}
as $\cV=\Ker J/\text{Im} E$.  We assume that $E$ is everywhere
injective and $J$ everywhere surjective on $M$, and that $\cV$  is a nonsingular,
stable holomorphic vector bundle.   These  are the geometric models
to which our methods apply. An intermediate step in the vanishing argument below involves setting $H=J=E=0$ in the M model -- we refer to this as the $O$ model.

The chiral superfield $S$ and the chiral Fermi multiplet $\Xi$ are the
``spectator" fields introduced in \cite{Distler:1995mi} to
maintain the K\"ahler parameters $r^a$ as RG invariant quantities. In
fact, the counterterm by which $r$ gets renormalized at one-loop is
proportional to the sum of the gauge charges of the scalar fields in
the theory.  In our models this is in general nonzero (this is related
to the fact that $V^+$ is not Calabi--Yau).  Introducing $S$ as above
cancels this.  The spectators earn their name because they interact
via a superpotential 
\begin{align}
\int d\theta^+ \Xi \,S\big|_{\thetab^+=0}  +\text{h.c.} ~.
\end{align}
This means these are massive fields and have no effect on the IR
dynamics of the theory, so one might question the relevance of
including them here. As we will see, in some cases accounting for
their presence allows the argument to proceed where it might otherwise
fail.  This is, of course, a technical matter.  If there are no
instanton corrections in the presence of spectators there are none in
their absence.  This demonstrates an important caveat to our work,
mentioned above.   When the argument of \cite{Beasley:2003fx} fails,
we cannot assert that instanton corrections {\sl do\/} destabilize the
model, only that this particular argument that they do not is
not valid.

\subsection*{Symmetries}

The action \eqref{eq:action1}-\eqref{eq:superpots} we have described
is invariant under (0,2) SUSY and the gauge symmetry $\GU(1)^R$, as
well as a global $\GU(1)_R\times \GU(1)_L$ symmetry acting as

\begin{align}
\xymatrix@R=1mm@C=3mm{
\text{fields}			&P^\alpha			&\Phi^i		&\Gamma^I		&\Lambda^A		&S			&\Xi			&\Sigma_\mu		&\Upsilon_a	\\
\GU(1)_L				&1				&0			&-1				&0				&1			&-1			&-1				&0		\\
\GU(1)_R				&1				&0			&0				&1				&1			&0			&1				&1
}
\end{align}
While $\GU(1)_R$ is believed to be the R-symmetry of the SCFT to which our
model is supposed to flow, the global $\GU(1)_L$ symmetry is
equally important for our purposes: in heterotic compactifications, it
can be used to construct a left-moving spectral flow operator and it provides a
linearly realized component of the space-time group.  The action
of global symmetries on charged fields is of course defined up to an
arbitrary action of the gauge symmetry generators.  We have here
chosen a representative action that is manifestly unbroken in the
classical vacua (when $r\in\cK_V$) comprising $M$.

These symmetries are respected by the classical action, but are in general
anomalous in the presence of non-trivial gauge fields. The
anomalies vanish when the charges satisfy
\begin{align}
\label{eq:anomalies}
d^a &= \sum_i q^a_i~, \nonumber\\
m^a &= \sum_I Q^a_I~, \nonumber\\
\sum_\alpha m^a_\alpha m^b_\alpha+\sum_i q_i^a q_i^b & = \sum_A d_A^ad_A^b + \sum_I Q_I^a Q_I^b~.
\end{align}
In terms of our geometric data, the first two conditions reflect the fact that 
\begin{align}
c_1(T_M)=c_1(\cV)=0~,
\end{align}
while the quadratic condition implies 
\begin{align}
\ch_2(T_M) = \ch_2(\cV)~.
\end{align}
Under these conditions, the $M$ model is believed to flow at low
energies to a nontrivial superconformal field theory which is in the
same moduli space as the nonlinear sigma model determined by the pair
$(M,\cV)$.  
Nonperturbative effects (worldsheet instantons) which can destroy
conformal invariance are captured by GLSM gauge instantons, which are
the subject of our investigation here.

\section{The argument}
\label{s:examples}

Let us first review the argument prescribed in \cite{Beasley:2003fx} for the vanishing of the instanton contributions 
to the superpotential $W$ for space-time gauge singlets in a (0,2) linear sigma model. 

The goal is to probe for a background space-time superpotential $W$.
A simple and direct way to achieve this is to compute the correlator
$\cC_{abc}=\la R_a R_b R_c \ra$, where $R_a$ is the vertex operator
representative for the K\"ahler modulus $\cR_a$ of $V$. In fact, for each instanton the
exponential factor $e^{I_0}$, where $I_0$ is the instanton classical action,
contains all the dependence on $\cR_a$ \cite{Dine:1986zy,Dine:1987bq}.  
The correlator $\cC_{abc}$ computes the third derivatives of $W$
with respect to $\cR_a$, 
thus it determines $W$ up to quadratic terms in the $\cR_a$. 
These terms are forbidden by standard $\alpha'$ non-renormalization theorems, hence $\cC_{abc}$ determines $W$ directly.

The computation is most easily done in the 
half-twisted model (see appendix \ref{app:twist}).  In this model, the
supercharge $\cQb_+$ becomes a nilpotent scalar symmetry generator, 
and correlators of $\cQb_+$-closed operators can be computed in its
cohomology.  On a genus-zero worldsheet, the twist can be realized by
spectral flow insertions and calculations in the twisted model produce
suitable correlators of the untwisted (physical) model.  

In order to determine the linear model representative of the
space-time mode $R_a$ we restrict our attention to the (0,2) gauge
multiplets. In fact, $R_a$ appears in the linear model through a F-I
term. Moreover, gauge singlets must have $q_L=0$ and bosonic vertex
operators have $q_R=1$. Finally,
\begin{align}
\cQb_+ \lambda_{a,-}&=0~, 	&\cQ_+\lambda_{a,-}&=\half\left(D_a-if_{a,01}\right)
\end{align}
determine $R_a=\lambda_{a,-}$.

The first step of the argument is to show that $\cC_{abc}$ vanishes in
the $O$ model. The idea is that the theory without
superpotential has a very large symmetry,
$G = \GU(1)^{\oplus(n+N+k+L+2)}$, where each matter superfield is rotated
separately, and the vertex operators $R_a$ are invariant under this.
This symmetry is generically broken by superpotential
couplings down to $\GU(1)_L$. If the zero-mode path
integral measure \footnote{We recall that the path integral for a correlator of $\cQb_+$-invariant operators
localizes on fixed loci of $\cQb_+$, given by zero-modes.}
in nontrivial topological sectors 
turns out not to be invariant under $G$, i.e.~the symmetry is
anomalous, then contributions to the invariant correlator $\cC_{abc}$ 
from this sector will vanish.  In practice, we follow
\cite{Beasley:2003fx}~and construct a $\GU(1)$ subgroup of $G$ that 
is rendered anomalous in {\sl all\/} topological sectors by the
twisting procedure, demonstrating that $\cC_{abc}=0$ identically in
the $O$ model. 

The second step uses the fact that $\cC_{abc}$ depends {\sl
  holomorphically\/} on $J,~H$ and $E$.  One can then examine the
contribution at arbitrary order in an expansion in these couplings.
If there is no term that can possibly absorb the fermion zero-modes in
the anomalous measure, then the correlator vanishes identically.  This
computation can be performed in each topological sector of the path
integral.

In the untwisted model, the limiting point $J=H=E=0$ is of course highly
singular.   Both $\sigma$ and $p$ acquire zero-modes and the space of
classical vacua is non-compact.   Such a singularity can invalidate
the order-by-order calculation described above.
The key observation of
\cite{Beasley:2003fx} is that in suitable examples these dangerous
zero-modes are absent in the half-twisted model.
For example, the bosons $\sigma$ always have zero-modes, but
in the twisted model (see appendix \ref{app:twist}) these fields
acquire a spin and their zero-modes are absent.   In general,
as we shall see below, $p$ zero-modes are not completely removed 
by the twisting.

Another approach was presented in \cite{Beasley:2003fx}, where the vanishing of $\cC_{abc}$ 
at any instanton number follows from an appropriate counting of fermi zero-modes
in the half-twisted model. This was applied in detail for heterotic compactifications described by half-linear sigma models,
but it also extends to linear models as well. However, as pointed out above, the same assumption of compactness is required
for this argument to be valid. For definiteness, we present our analysis 
of the linear model following the approach of \cite{Beasley:2003fx} reviewed above, as 
our results will not depend on this choice.

\subsection{The quintic}
\label{ss:quintic}

Let us review how all of this works for the linear model describing the deformations of the tangent bundle $T_M$ over the 
quintic hypersurface $M$ in $V=\P^4$. The gauge charges for the (2,2)
multiplets $\Phi^i=(\Phi^i,\Gamma^i)$ and $P=(P,\Lambda)$ are 
\begin{align}
\xymatrix@R=1mm@C=2mm{
\text{fields} 	&P		&\Phi^1	&\Phi^2	&\Phi^3	&\Phi^4	&\Phi^5	\\
\GU(1)		&-5		&1		&1		&1		&1		&1	 
}
\end{align}
The K\"ahler cone $\cK_V=\cK_c$ here is simply given by $r\geq0$ and the relevant instantons are defined by $\cK_V^\lor=\{n\geq0\}$.
The $O$ model has as a target space the total space of the anticanonical bundle on $V$, $\text{tot}\left(\cO(-5)\rightarrow \P^4  \right)$. 
First of all, we check that the moduli space of gauge instantons for this model is compact. Indeed, we verify that there are no holomorphic sections of
\begin{align}
p \quad \leftrightarrow \quad \Gamma(K^\half\otimes \cO(-5n))~,
\end{align}
and thus $p$ has no zero-modes. This, together with the fact that there
are no zero-modes of $\sigma$ shows that the space of zero-modes is
compact in any topological sector.
Next, by looking at the 
degree of the line bundles of the half-twisted model in \eqref{eq:twistmatter} we see that the relevant fermions zero-modes are
\begin{align}
\xymatrix@R=1mm@C=4mm{
\text{fields	}	&\psib^i		&\gamma^i	&\etab				&\chi			\\
\text{bundle}	&\cOb(n)		&\cO(n-1)		&\cO(5n)				&\cOb(5n-1)	\\
\text{\# z.m.}	&n+1		&n			&5n+1				&5n
}
\end{align}
The fermion contribution to the zero-mode path-integral measure is then given by
\begin{align}
d\mu_F = d\lambdab_{-} d\etab d\chi \prod_i d\psib^id\gamma^i~.
\end{align}
Now, the $O$ model is invariant under a symmetry $\GU(1)_C$ which assigns charge $+1$ to the multiplets $\Phi^i$ and
leaves everything else invariant. Under this symmetry the measure above transforms with charge $+5$. Hence, the correlator $\cC$
vanishes in the $O$~model. 
The holomorphic superpotential couplings are given by
\begin{align}
\cL_{\text{Yuk}}\big|_{\Jb=\Hb=\Eb=0} &=  -\gammab^i E^i \lambda_{+} + \gamma^iJ_{i}\chi + \eta H_{,j}\psi^j~,
\end{align}
where $H$ is a quintic polynomial defining the hypersurface $M$, $J_i$ are generic quartic polynomials and $E^i$ are generic linear polynomials
subject to \eqref{eq:SUSYcond}.
Clearly, each coupling transforms under $\GU(1)_C$ with either charge +5 or is neutral. By the argument above the correlator 
$\cC$ vanishes in the full theory and there are no instanton corrections to the space-time superpotential.

\subsection{A counter-example}
\label{ss:counterex}

Let us consider a two-parameter model with the following charge assignments
\begin{align}
\xymatrix@R=1mm@C=2mm{
\text{fields}	&\Phi^{1,2,3}	&\Phi^{4,5}	&\Phi^{6,7}	&\Lambda^1	&\Lambda^2	&\Gamma^{1,2}	&\Gamma^3	&\Gamma^{4,5,6,7}		&P^1	&P^2		\\
\GU(1)_1		&1			&1			&0			&-3			&-2			&2				&1			&0					&-4		&-1			\\
\GU(1)_2		&1			&0			&1			&-3			&-2			&0				&1			&1					&-2		&-3		
}
\end{align}
In the geometric phase it describes a complete intersection $M$ of degree $(3,3)$ and $(2,2)$ hypersurfaces in the toric variety $V$ defined by the charges
\begin{align}
\begin{pmatrix}
1&1&1&1&1&0&0\\
1&1&1&0&0&1&1
\end{pmatrix}~.
\end{align}
It is useful to write the maps defining the superpotential more explicitly. For ease of notation, let us denote $x=\{\phi^{1,2,3}\},~ y=\{\phi^{4,5}\},~z=\{\phi^{6,7}\}$, 
as well as $\Gamma^{(1)}=\{\Gamma^{1,2}\},~
\Gamma^{(2)}=\{\Gamma^{3}\},~\Gamma^{(3)}=\{\Gamma^{4,5,6,7}\}$, and a
condensed notation in which, e.g.~$x^k$ denotes a generic homogeneous
polynomial of degree $k$ in $\phi^{1,2,3}$.
With this notation the maps are given as
\begin{align}
\label{eq:exJs}
J_{(1)}&=p^1(x^2+xyz+y^2z^2)~, \nonumber\\
J_{(2)}&=p^1(xy^2+y^3z)+p^2z^2~,\nonumber\\
J_{(3)}&=p^1(xy^3+y^4z)+p^2(xz+yz^2)~,
\end{align}
while the equations defining the complete intersections are
\begin{align}
\label{eq:exHs}
H_1 &= x^3 + x^2yz+xy^2z^2+y^3z^3~,\nonumber\\
H_2 &= x^2 + xyz+y^2z^2~.
\end{align}
The complete intersection $M$ is realized in the cone $\cK_V=\{r_1>0,r_1-r_2<0\}$, where
the irrelevant ideal is $B=(xy)(z)$. Since the $z$'s are not both allowed to vanish and the coefficients
in the expressions above are generic, we have that 
\begin{align}
&\underbrace{(xz+yz^2)}_{4 \text{ of these}} = x^3 + x^2yz+xy^2z^2+y^3z^3 = x^2 + xyz+y^2z^2=0~,\nonumber\\
&\underbrace{(x^2+xyz+y^2z^2)}_{3 \text{ of these}} = xy^2+y^3z = \underbrace{(xy^3+y^4z)}_{4 \text{ of these}}= x^3 + x^2yz+xy^2z^2+y^3z^3 =0~,
\end{align}
have no solutions compatible with the ideal $B$. Thus $p^1=p^2=0$ and there are no flat directions in this phase.

Now, note that $-p^2\in \cK_V$ but $-p^1\notin\cK_V$. Therefore there are instantons contributing for this phase for which $p^1$ develops zero-modes in the $O$~model. 
We can see this explicitly. 
Gauge instantons in this model have instanton numbers
$n_a\in\cK_V^\lor$,
i.e.~$n_2>0$ and $n_1+n_2>0$. From appendix \ref{app:twist} we see that the zero-modes of $p^1$ are in one to one correspondence with
holomorphic section of the bundle 
\begin{align}
p^1 \quad \leftrightarrow \quad \Gamma(K^\half\otimes \cO(-4n_1-2n_2))~,
\end{align}
and the number of such sections is non-zero when $2n_1+n_2<0$. The subcone defined by $(2n_1+n_2<0)\cap \cK_V^\lor$ is non empty,
and the moduli space of gauge instantons of the $O$ model is not
compact.  The twisted $O$~model calculation in these sectors is
ill-defined and the argument from holomorphy does not exclude 
instanton corrections to $\cC_{abc}$.

\section{The vanishing theorem}
\label{s:theorem}

The example of the previous section shows that a generic (0,2) GLSM is not protected 
from worldsheet instanton corrections. In this section we undertake the task of constructing
a class of models for which the vanishing theorem holds.
In fact, a necessary condition for the vanishing argument to apply is that there exists 
a cone $\cK_V\subseteq\cK_c$ such that
\begin{enumerate}
\item the M model defined in $\cK_V$ is nonsingular;
\item the $O$ model of the half-twisted theory has a compact moduli
  space of gauge instantons for any $n_a\in\cK_V^\lor$.
\end{enumerate}
Notice that as advertised above, due to the twist the bosons $\sigma$ acquire a spin and do not have zero-modes.
A quick inspection at the form of the $E$-couplings \eqref{eq:Ecoup1} and \eqref{eq:Ecoup2} 
implies that setting $E=0$ does not lead to any singularities in the half-twisted theory. 

\subsection{$O$ model gauge instanton moduli space}
\label{ss:compact}

While the moduli space of gauge instantons for the $V$ model is compact, as we have seen above, 
there can be unbounded zero-modes coming from the $p^\alpha$ fields. This occurs when, for a given subcone $\cK_V\subseteq\cK_c$
we have $m_\alpha \notin \cK_V$ for some $\alpha$.
Hence, a necessary condition for the argument to work is that there exists a nonsingular subcone $\cK_V\subseteq\cK_c$ 
such that $m_\alpha \in \cK_V$~$\forall\alpha$.

The discussion so far did not take into account the spectator boson $s$, whose expectation value is set to zero, and whose
zero-modes could also be fatal for our assumption of compactness.
In order to establish when this is the case, we need the following simple fact:
the cone $\widehat\cK_c$ defined by adjoining the vector $m-d$ to $\cK_c$ is convex unless $d-m \in \cK_c$.
In fact, 
$\widehat\cK_c$ fails to be convex if we can write
\begin{align}
\sum_i \alpha_i q_i + \beta (m-d)=0~,
\end{align}
with $\alpha_i,\beta\geq 0$ and not all vanishing.  Because $\cK_c$ is convex by assumption, we must have 
$\beta$ strictly positive.  This means $\beta(d-m) = \sum_i\alpha_i q_i$, 
i.e.~$(d-m)\in\cK_c$.

This little result suggests there are three separate cases we should consider:
\begin{enumerate}
\item $d-m \in \cK_V$. By \eqref{eq:twistmatter} $s$ has no zero-modes.  In fact, by looking at the degree 
\begin{align}
d_S = (m^a-d^a)n_a,
\end{align}
we have $d_S\leq0$ $\forall n\in \cK_V^\lor$.
\item $d-m \in \cK_c$ but $d-m\notin\cK_V$. In this case there exist $n\in \cK_V^\lor$ such that $d_S>0$ and $s$ has zero-modes. 
The half-twisted $O$ model develops $s$-flat directions and is therefore singular. 
\item $d-m \notin \cK_c$.  In this case $s$ always has zero-modes, but 
by the result above, together with the fact that a toric variety is compact if and only if the geometric cone is strongly convex,
 the moduli space of instantons for the $O$ model is nevertheless compact.
\end{enumerate}
We can now summarize the set of conditions we are going to assume for our vanishing theorem:
there exists a nonsingular subcone $\cK_V\subseteq \cK_c$ such that 
the gauge charge vectors for the fields $p^\alpha$ and $s$ satisfy
\begin{align}
&m_\alpha \in \cK_V\quad \forall\alpha~, &d-m&\in \cK_V\quad \text{ or } \quad d-m \notin \cK_c~. 
\end{align}

\subsection{A classical symmetry}

For the remaining of this section we restrict our attention to models obeying the conditions above.
To proceed with the argument we need to construct a suitable $\GU(1)_C$ subgroup
of the symmetry group of the $O$ model. Let us choose the charges 
for the matter fields under this ``classical'' symmetry as
\begin{align}
\xymatrix@R=2mm@C=4mm{
\text{fields}			&P^\alpha		&\Phi^i		&\Gamma^I		&\Lambda^A		&S			&\Xi	\\
\GU(1)_C				&0			&q_i^C		&Q^C_I			&0				& q^C_S		& Q^C_\Xi
}
\end{align}
while the gauge fields are  invariant.
This symmetry will be non-anomalous (before twisting) if
\begin{align}
\label{eq:classsyman}
 \sum_i q_i^a q_i^C + (m^a-d^a)q^C_S = \sum_I Q_I^a Q^C_I + (d^a-m^a)Q^C_\Xi~,
\end{align}
for $a=1,\dots,R$.

\subsection*{The measure}

First, let us look at the zero-mode contribution to the path integral measure. 
In particular, we are going to focus only on the fermionic part of the measure. In fact, the
form of the maps \eqref{eq:twistmatter}, together with our assumption of compactness yield an exact balance 
between holomorphic and anti-holomorphic bosonic zero-modes.
It is convenient to write the fermionic measure as
\begin{align}
d\mu_F = d\mu_G d\mu_M d\mu_S~,
\end{align}
where the three factors correspond to the measure for the gauge, matter and spectator fields respectively.
From \eqref{eq:twistgauge} it follows that the gauge measure is simply given by
\begin{align}
d\mu_G = \prod_a d\lambdab_{-,a}~.
\end{align}
For the matter fields we have
\begin{align}
\xymatrix@R=2mm@C=5mm{
\text{fields}			&\chi^\alpha						&\chib^\alpha						&\psi^i						&\psib^i		\\
\text{bundle}			&\Kb^\half \otimes \cOb(-d_\alpha)		&\Kb^\half\otimes\cOb(d_\alpha)		&\Kb\otimes\cOb\left(-d_i \right)		&\cOb\left(d_i \right)		\\
\text{\# z.m.}			&\max(0,-d_\alpha)=-d_\alpha			&\max(0,d_\alpha)=0				&\max(0,-d_i-1)					&\max(0,d_i+1)
}
\end{align}
where we used the fact that $m_\alpha\in\cK_V$ implies $d_\alpha\leq0$, as well as
\begin{align}
\xymatrix@R=2mm@C=5mm{
\text{fields}			&\gamma^I						&\gammab^I						&\eta^A						&\etab^A				\\
\text{bundle}			&K^\half \otimes \cO\left(D_I \right)		&K^\half\otimes\cO\left(-D_I\right)		&K\otimes\cO\left(D_A \right)		&\cO\left(-D_A \right)		\\
\text{\# z.m.}			&\max(0,D_I)						&\max(0,-D_I)						&\max(0,D_A-1)				&\max(0,-D_A+1)
}
\end{align}
The matter measure then reads
\begin{align}
 d\mu_M =\prod_\alpha d\chi^\alpha \prod_{i|d_i\geq0} d\psib^i \prod_{i|d_i<0} d\psi^i \prod_{I|D_I\geq0}d\gamma^I \prod_{I|D_I<0}d\gammab^I \prod_{A|D_A>0}d\eta^A \prod_{A|D_A\leq0}d\etab^A~,
\end{align}
and it is easy to check that it is gauge-invariant.

Finally, for the spectators
\begin{align}
\xymatrix@R=2mm@C=5mm{
\text{fields}			&\xi_+							&\xib_+							&\xi_-						&\xib_-				\\
\text{bundle}			&\Kb^\half \otimes \cOb\left(-d_S \right)	&\Kb^\half\otimes\cOb\left(d_S\right)		&K^\half\otimes\cO(-d_S)			&K^\half\otimes\cO(d_S)		\\
\text{\# z.m.}			&\max(0,-d_S)						&\max(0,d_S)						&\max(0,-d_S)					&\max(0,d_S)
}
\end{align}
Here we need to distinguish two cases, according to whether $d-m\in\cK_V$ or $d-m\notin\cK_c$, and we obtain
\begin{align}
\label{eq:specmes}
d\mu_S = 
\begin{cases} 
d\xi_+d\xi_-~~  \text{ if } d_S<0~,\\
d\xib_+d\xib_-~~ \text{ if } d_S>0~.
\end{cases}
\end{align}
Of course, if $d_S=0$ we simply ignore this factor.

Now we can finally determine how the measure transforms under the symmetry $\GU(1)_C$ defined above.
The gauge measure is invariant, while for the matter factor we obtain 
\begin{align}
 q^C(d\mu_M)&= \sum_{i|d_i\geq0}(d_i+1)q^C_i + \sum_{i|d_i<0}(-d_i-1)(-q^C_i) + \sum_{I|D_I\geq0}D_I(-Q^C_I) + \sum_{I|D_I<0}(-D_I)Q^C_I ~.    
 \end{align}
Finally, for the spectator measure in both cases of \eqref{eq:specmes} we get
\begin{align}
q^C(d\mu_S)=d_S(q^C_S+Q^C_\Xi)~,
\end{align}
Let us observe at this point that a very simple solution to \eqref{eq:classsyman} is given by
\begin{align}
 q^C_i=Q^C_I=q^C_S=1~, \qquad Q^C_\Xi=0~,
\end{align}
where it is easy to verify that the equality holds by \eqref{eq:anomalies}.
Plugging these values into the expressions above we find that the total fermionic zero-mode measure in the twisted model transforms with charge $q^C(d\mu_F)=n$,
where we recall that $n$ is the number of one-dimensional cones of the fan $\Delta_V$ for the toric variety $V$, and in particular is strictly positive.
Thus, the fermion zero-modes cause $\GU(1)_C$ to be anomalous, and $\cC_{abc}$ vanishes in the $O$ model.

\subsection*{The superpotential couplings}

Let us turn to the analysis of the superpotential couplings in the action. The relevant Yukawa couplings are 
\begin{align}
\label{eq:yuk}
\cL_{\text{Yuk}}\big|_{\Jb=\Hb=\Eb=0} &=  -\gammab^IE^I_{,\mu}\lambda_{\mu,+} 
  + \gamma^IJ_{I\alpha}\chi^\alpha + \eta^A H_{A,j}\psi^j~,
\end{align}
 where we have set $\sigma_\mu=p^\alpha=0$, as they have no zero-modes.
 We immediately see that all couplings, when non-zero, have the following lower bounds on the charges
 \begin{align}
\xymatrix@R=2mm@C=5mm{
\text{couplings}			&\gammab^IE^I_{,\mu}\lambda_{\mu,+} 	&\gamma^IJ_{I\alpha}\chi^\alpha	&\eta^A H_{A,j}\psi^j			\\
\GU(1)_C				&\geq0							&\geq2						&\geq1	
}
\end{align}
 In particular, we note that these values are all non-negative and therefore it is not possible
 to absorb the zero-modes in excess in the measure by bringing down fermion terms from the action. The correlator
 $\cC_{abc}$ thus vanishes at all orders in the superpotential couplings, which concludes the proof that instantons do not contribute
 to the space-time superpotential in our class of models.
 
 Note that we ignored the anti-holomorphic functions $\Jb,~\Hb$ and $\Eb$ in \eqref{eq:yuk}.
 This is in fact legitimate since, as observed above, half-twisted correlators of $\cQb_+$-closed operators have a holomorphic dependence on $J,~H$ and $E$.

\section{Outlook}
\label{s:conclusions}

In this work, we investigated the details of the elegant argument of
\cite{Beasley:2003fx} for the absence of instanton
corrections to the space-time superpotential in heterotic
compactifications based on (0,2) GLSMs.  We have not been able to
extend the argument to the most general case.

The immediate question raised is:  are some of these vacua in fact
destabilized by instantons?  One clear way to resolve this would be 
to produce an argument that holds in more 
generality.  It is possible, however, that no such argument can be
found and that in fact instanton corrections do arise.   One way
to detect such corrections would be an indirect approach, in 
which properties of the solution, such as the dimension of the 
space of massless gauge-neutral scalar fields, 
are compared at different limiting points in the moduli space.
A more direct approach would be to compute the instanton contributions
explicitly.  Perhaps the GLSM can provide a framework 
within which these calculations, which have proved difficult in 
general, are tractable.

On the other hand, we have now an extensive class of (0,2) models
which are truly conformally invariant.  These can be used to explore
the moduli space of (0,2) theories without a (2,2) locus, extending
recent work that has focused on deformations of (2,2) models
\cite{McOrist:2008ji,Aspinwall:2014ava,Donagi:2014koa}.  In
particular, one could look for special loci, e.g. good hybrid models
\cite{Bertolini:2013xga} or Landau-Ginzburg points and hope to learn
something about the structure of the resulting theories. In
particular, hybrid models could be a promising laboratory for
explicit computations of worldsheet instantons, given the simpler
structure of rational curves on the lower dimensional base instead
than on a CY three-fold.

Recently it has been shown that other ``bad" things can happen in
(0,2) models \cite{Bertolini:2014ela}.  In particular, it is shown, in the context of Landau-Ginzburg models, 
that the common assumption that accidental IR symmetries do not spoil the correspondence between operators in the IR and 
the ones in the UV is not guaranteed in (0,2) models. 
When this occurs, the structure of the conformal manifold is dramatically modified.
There is a priori no
reason that would prevent the same phenomenon from happening in a
generic phase of a GLSM.  For example, one could realize
one of the ``accidental" LG theories as a phase of a GLSM and study
how this pathology is realized in the geometric phase.  This could
shed new light on the conditions for the data $(M,\cV)$ to lead to
consistent heterotic backgrounds.

\appendix

\section{Linear model conventions}
\label{app:linearmodel}

\subsection{(0,2) superspace}

We work in (0,2) superspace\footnote{More details may be found in \cite{Wess:1992cp}.} with coordinate $x^{\pm},\theta^+,\thetab^+$. 
The supercharges are given by
\begin{align}
\cQ_+ &=  {\p \over \p \theta^+ } +i\thetab^+ \nabla_+~, &\cQb_+ &= - {\p \over \p \thetab^+ } -i\theta^+ \nabla_+~,
\end{align}
where $\p_+ = \p/\p x^+$ and $\nabla_+$ is the covariant gauge derivative. We also have the superderivatives 
\begin{align}
\cD_+ &=  {\p \over \p \theta^+ } -i\thetab^+ \nabla_+~, &\cDb_+ &= - {\p \over \p \thetab^+ } +i\theta^+ \nabla_+~.
\end{align}
The non-trivial anti-commutation relations are
\begin{align}
\AC{\cQ_+}{\cQb_+} &= -2i\nabla_+~, 	&\AC{\cD_+}{\cDb_+}&=2i\nabla_+~.
\end{align}

\subsection{Field content}

There are two types of multiplets in the (0,2) models we consider in this work.
\begin{enumerate}
\item \underline{Gauge fields multiplets}. We have
\begin{align}
\label{eq:gaugefields}
V_{a,-}&=v_{a,-}-2i\theta^+\lambdab_{a,-}-2i\thetab^+\lambda_{a,-}+2\theta^+\thetab^+ D_a~,\nonumber\\
\Sigma_\mu &= \sigma_\mu + \sqrt2\theta^+ \lambda_{\mu,+}-i\theta^+\thetab^+\p_+ \sigma_\mu~,
\end{align}
where $a=1,\dots,R$ and $\mu=1,\dots,m$.
The multiplets $\Sigma_\mu$ are neutral chiral multiplets which in (2,2) theories 
\footnote{In (2,2) theories we have $R=m$.} combine
with the (0,2) gauge multiplets into (2,2) gauge multiplets.  The twisted chiral gauge invariant field strength 
is defined as
\begin{align}
\Upsilon_a &= \CO{\cDb_+}{\nabla_-} \nonumber\\
&= i\cDb_+ V_{a,-} + \theta^+ \nabla_- v_{a,+}  \nonumber\\
&=-2\lambda_{a,-} -i\theta^+ (D_a-if_{a,01})-i\theta^+\thetab^+ \p_+\lambda_{a,-}~. 
\end{align}
\item \underline{Matter multiplets}. Here we have bosonic chiral (anti-chiral) multiplets
\begin{align}
P^\alpha&=p^\alpha+\sqrt2 \theta^+ \chi^\alpha - i\theta^+\thetab^+ \nabla_+ p^\alpha~,
&\Pb^\alpha&=\pbb^\alpha-\sqrt2 \thetab^+ \chib^\alpha + i\theta^+\thetab^+ \nabla_+ \pbb^\alpha~, \nonumber\\
\Phi^i&=\phi^i+\sqrt2 \theta^+ \psi^i - i\theta^+\thetab^+ \nabla_+ \phi^i~,
&\Phib^i&=\phib^i-\sqrt2 \thetab^+ \psib^i + i\theta^+\thetab^+ \nabla_+ \phib^i~, \nonumber\\
S &=s+\sqrt2 \theta^+ \xi_+ - i\theta^+\thetab^+ \nabla_+ s ~,
&\Sb &=\sb-\sqrt2 \thetab^+ \xib_+ + i\theta^+\thetab^+ \nabla_+ \sb ~,
\end{align}
where $\alpha=1,\dots,k$ and $i=1,\dots,n$. We also have fermionic matter multiplets, which we again divide into three groups 
\begin{align}
\Gamma^I &= \gamma^I - \sqrt2 \theta^+ G^I - i\theta^+\thetab^+ \nabla_+ \gamma^I -\sqrt2 \thetab^+ E^I(\Phi,\Sigma)~, \nonumber\\
\Lambda^A &= \eta^I - \sqrt2 \theta^+ F^A - i\theta^+\thetab^+ \nabla_+ \eta^A -\sqrt2 \thetab^+ E^A(P,\Phi,\Sigma)~, \nonumber\\ 
\Xi &= \xi_- - \sqrt2 \theta^+ K - i\theta^+\thetab^+ \nabla_+ \xi_-~,
\end{align}
as well as their complex conjugate 
\begin{align}
\Gammab^I &= \gammab^I - \sqrt2 \thetab^+ \Gb^I + i\theta^+\thetab^+ \nabla_+ \gammab^I -\sqrt2 \theta^+ \Eb^I(\Phib,\Sigmab)~, \nonumber\\
\Lambdab^A &= \etab^I - \sqrt2 \thetab^+ \Fb^A + i\theta^+\thetab^+ \nabla_+ \etab^A -\sqrt2 \theta^+ \Eb^A(\Pb,\Phib,\Sigmab)~, \nonumber\\ 
\Xib &= \xib_- - \sqrt2 \thetab^+ \Kb + i\theta^+\thetab^+ \nabla_+ \xib_- ~.
\end{align}
Here the indices are $I=1\dots,N$ and $A=1,\dots,L$.
\end{enumerate}

\subsection{The action}

Let us list the various terms that appear in the action for the (0,2) linear models we consider in this work. 
We have the kinetic term for the gauge fields\footnote{For simplicity, we have set equal all gauge coupling constants.}
\begin{align}
\label{eq:action1}
\cL_{G,K} = {1\over8e^2} \int d^2\theta^+ \Tr \Upsilonb_a\Upsilon_a = {1\over 2e^2}\left[ 2i\lambdab_{a,-}\p_+ \lambda_{a,-} + D_a^2 + f_{a,01}^2 \right]~, 
\end{align}
as well as the kinetic term for the $\Sigma_\mu$ fields
\begin{align}
\cL_{\Sigma,K} = {i\over2e^2} \int d^2\theta^+ \Sigma_\mu\nabla_-\Sigma_\mu = {1\over e^2}\left[ \p_+\sigmab_\mu\p_-\sigma_\mu + i\lambdab_{\mu,+}\p_- \lambda_{\mu,+}  \right]~.
\end{align}
Then we have the kinetic terms for the various matter fields. These are given as
\begin{align}
\cL_{\Phi,K} = {i\over2} \int d^2\theta^+ \Phib^i \nabla_- \Phi^i 
&= \half \left(\nabla_+\phib^i\nabla_- \phi^i + \nabla_-\phib^i\nabla_+\phi^i\right) + i\psib^i \nabla_- \psi^i \nonumber\\
&\quad+ i\sqrt2 q_i^a \left(\psib^i\lambdab_{a,-}\phi^i -\phib^i \lambda_{a,-} \psi^i\right) +q_i^aD_a \phib^i\phi^i ~,  \nonumber\\
\cL_{\Gamma,K} = {1\over2} \int d^2\theta^+ \Gammab^I \Gamma^I 
&= i\gammab^I \nabla_+ \gamma^I + \Gb^I G^I - \Eb^IE^I  \nonumber\\
&\quad -\gammab^I E^I_{,j}\psi^j - \gammab^IE^I_{,\mu}\lambda_{\mu,+} - \Eb^I_{,j}\psib^j\gamma^I - \Eb^I_{,\mu}\lambdab_{\mu,+}\gamma^I~,
\end{align}
and similarly
\begin{align}
\cL_{P,K} &= {i\over2} \int d^2\theta^+ \Pb^\alpha \nabla_- P^\alpha
= \half \left(\nabla_+\pbb^\alpha\nabla_- p^\alpha + \nabla_-\pbb^\alpha\nabla_+p^\alpha\right) + i\chib^\alpha \nabla_- \chi^\alpha  \nonumber\\
&\qquad \qquad \qquad \quad\qquad\qquad 
- i\sqrt2 m_\alpha^a\left( \chib^\alpha\lambdab_{a,-} p^\alpha -  \pbb^\alpha\lambda_{a,-} \chi^\alpha\right) -m_\alpha^a D_a \pbb^\alpha p^\alpha~,  \nonumber\\
\cL_{\Lambda,K} &= {1\over2} \int d^2\theta^+ \Lambdab^A \Lambda^A
= i\etab^A \nabla_+ \eta^A + \Fb^A F^A - \Eb^AE^A \nonumber\\
&\qquad \qquad \quad\qquad\qquad-\etab^A E^A_{,j}\psi^j - \etab^A E^A_{,\mu}\lambda_{\mu,+} - \Eb^A_{,j}\psib^j\eta^A - \Eb^A_{,\mu}\lambdab_{\mu,+}\eta^A~, \nonumber\\
\cL_{S,K} &= {i\over2} \int d^2\theta^+ \Sb \nabla_- S 
= \half \left(\nabla_+\sb\nabla_- s + \nabla_-\sb \nabla_+s\right) + i\xib_+ \nabla_- \xi_+ \nonumber\\
&\qquad \qquad \qquad \quad \qquad + i\sqrt2 (m^a-d^a)\left( \xib_+ \lambdab_{a,-} s  -\lambda_{a,-} \xi_+\sb \right)+(m^a-d^a) D_a \sb s ~,  \nonumber\\
\cL_{\Xi,K} &= {1\over2} \int d^2\theta^+ \Xib \Xi
= i\xib_- \nabla_+ \xi_- + \Kb K ~.
\end{align}
The Fayet-Iliopoulos terms action arises as a linear twisted superpotential for the twisted chiral fields $\Upsilon_a$
\begin{align}
\cL_{\text{F-I}} &= \ff14 \int d\theta^+  \Upsilon_a \tau^a\big|_{\thetab^+=0}  +\text{h.c.} = -D_a r^a + {\theta^a\over 2\pi} f_{a,01}~,
\end{align}
where $\tau^a=ir^a + \theta^a/2\pi$ are the complexified F-I parameters.
Finally, the matter superpotential is a sum of three terms
\begin{align}
\label{eq:superpots}
\cL_{J}&=-{1\over\sqrt2}\int d\theta^+  \Gamma^I J_I(P,\Phi)\big|_{\thetab^+=0}  +\text{h.c.}
= G^I p^\alpha J_{I\alpha} + \gamma^Ip^\alpha J_{I\alpha,j}\psi^j +\gamma^IJ_{I\alpha}\chi^\alpha + \text{h.c.}~, \nonumber\\
\cL_{H}&=-{1\over\sqrt2}\int d\theta^+  \Lambda^A H_I(\Phi)\big|_{\thetab^+=0}  +\text{h.c.} 
= F^A H_{A} + \eta^A  H_{A,j}\psi^j+ \text{h.c.}~,\nonumber\\
\cL_{S}&=-{1\over\sqrt2}\int d\theta^+  \Xi S \big|_{\thetab^+=0}  +\text{h.c.} 
= Ks + \xi_- \xi_+ + \text{h.c.}~.
\end{align}
The last term explicitly shows that all the excitations of the spectator fields are massive and they do not affect the low energy physics.
In \eqref{eq:superpots} we implemented the form for the superpotential \eqref{eq:Jsup}.

\section{The half-twist}
\label{app:twist}

In order to probe for a background space-time superpotential $W$ it is convenient to half-twist the model, that is we twist by $J_H=J_R/2$, where 
$J_R$ is the generator of the right-moving R-symmetry. 
We implement this by redefining the Lorentz generator $J_L$ as
\begin{align}
J'_L= J_L - J_R/2~.
\end{align}
Explicitly, for the gauge fields we have
\begin{align}
\label{eq:twispinsgauge}
\xymatrix@R=1mm@C=4mm{
\mbox{fields} 	& \sigma_\mu 	& \sigmab_\mu	& \lambda_{+,\mu}	& \lambdab_{+,\mu}	&\lambda_{-,a}  	&\lambdab_{-,a} 	\\
J_L		 	& 0 			& 0 			& \half		   	& \half 			& -\half 			& -\half   		     		\\
J'_L			& -\half 		& \half		& \half 		   	& \half 			& -1 				& 0 		      	}
\end{align}
while for the matter fields we have instead
\begin{align}
\label{eq:twispinsmatter1}
\xymatrix@R=1mm@C=4mm{
\mbox{fields} 	&p^\alpha 	&\pbb^\alpha	&\phi^i	&\phib^i	&\chi^\alpha	&\chib^\alpha	&\psi^i	&\psib^i	\\
J_L		 	&0 			&0 			&0		&0	 	&\half		&\half		&\half	&\half     	\\
J'_L			&-\half		&\half		&0 		&0 		&\half		&\half		&1		&0		}
\end{align}
and
\begin{align}
\label{eq:twispinsmatter2}
\xymatrix@R=1mm@C=3mm{
\mbox{fields} 	&\eta^A		&\etab^A		&\gamma^I	&\gammab^I	&s		&\sb		&\xi_+	&\xib_+	&\xi_-	&\xib_-	\\
J_L		 	&-\half 		&-\half  		&-\half 		&-\half  		&0		&0		&\half	&\half	&-\half	&-\half	\\
J'_L			&-1 			&0			&-\half		&-\half		&-\half	&\half	&\half	&\half	&-\half	&-\half     	}
\end{align}
In the twisted model the supercharge $\cQb_+$ becomes a worldsheet
scalar.  $\cQb_+$-exact operators will decouple from the correlators
of $\cQb_+$-closed fields, to which we restrict our attention.  In
particular, the kinetic terms for all fields are $\cQb_+$-exact up to
a topological term determined by the gauge bundle on the world-sheet 
$\Sigma=\P^1$ via the instanton numbers 
\begin{align}
n_a=-{1\over2\pi}\int f_{a,01}~.
\end{align}
The integral over field configurations breaks up into a sum over topological
sectors indexed by $n_a$.   For $r\in \cK_V$, these lie in
$\cK_V^\lor$, and the classical action weights the contribution of
each sector by $\prod_a q_a^{n_a}$ where $q_a = e^{-2\pi r_a + i\theta_a}$.
Extracting this topological contribution we can 
perform the computation within each topological sector semiclassically, and the path
integral reduces to an integral over the zero modes of the fields.  

The space of zero modes to which the path integral reduces in each
sector can be represented as the space of (anti-) holomorphic sections
of appropriate line bundles over $\Sigma$.  
Explicitly, the gauge fields take values in 
\begin{align}
\label{eq:twistgauge}
\xymatrix@R=0mm@C=4mm{
\sigma_a			&\leftrightarrow 	&K^{\half}				&&		&\sigmab_a		&\leftrightarrow	&\Kb^{\half}  \\
\lambda_{+,\mu} 	&\leftrightarrow		&\Kb^{\half}			&&		&\lambdab_{+,\mu}  	&\leftrightarrow&  \Kb^{\half}  \\
\lambda_{-,a}  		&\leftrightarrow		&K 					&&		&\lambdab_{-,a}	&\leftrightarrow&  \cO 	 
}
\end{align}
where $K = \cO(-2)$ is the canonical bundle.  
For the matter fields we have instead
\begin{align}
\label{eq:twistmatter}
\xymatrix@R=0mm@C=4mm{
p^\alpha		&\leftrightarrow 		& K^\half\otimes\cO(d_\alpha)			&&	&\pbb^\alpha		&\leftrightarrow 	& \Kb^\half\otimes\cOb(d_\alpha)		\\								
\phi^i		&\leftrightarrow 		& \cO(d_i)							&&	&\phi^i			&\leftrightarrow 	& \cOb(d_i)					\\
\psi^i			&\leftrightarrow 		& \Kb \otimes\cOb(-d_i)				&&	&\psib^i			&\leftrightarrow 	& \cOb(d_i)			  \\		
\gamma^I 		&\leftrightarrow 		&  K^\half\otimes\cO(D_I)				&&	&\gammab^I		&\leftrightarrow 	& K^\half\otimes\cO(-D_I)			  \\		   
\eta^A	 	&\leftrightarrow 		&  K\otimes\cO(D_A)					&&	&\etab^A			&\leftrightarrow 	& \cO(-D_A)				  \\	
\chi^\alpha 	&\leftrightarrow 		&  \Kb^\half\otimes\cOb(-d_\alpha)		&&	&\chib^\alpha		&\leftrightarrow 	& \Kb^\half\otimes\cOb(d_\alpha)			  \\	
s			&\leftrightarrow 		&K^\half\otimes\cO\left(d_S \right)		&&	&\overline{s}		&\leftrightarrow 	&\Kb^\half\otimes\cOb\left(d_S \right)			\\
\xi_+			&\leftrightarrow 		&\Kb^\half\otimes\cOb\left(-d_S \right)	&&	&\xib_+			&\leftrightarrow 	&\Kb^\half\otimes\cOb\left(d_S \right)		\\
\xi_-			&\leftrightarrow 		&K^\half\otimes\cO\left(-d_S \right)		&&	&\xib_-			&\leftrightarrow 	&K^\half\otimes\cO\left(d_S \right)	
	}
\end{align}
where the various degrees are defined as 
\begin{align}
d_\alpha &= -m^a_\alpha n_a~, &d_i&=q_i^an_a ~, &D_I&=Q^a_I n_a~, &D_A&= -d_A^a n_a~, &d_S&=(m^a-d^a)n_a~.
\end{align}
Note that it turned out to be convenient to use a hermitian metric on the appropriate bundles on $\P^1$ to redefine some of the fields \cite{Katz:2004nn}.
By examining the half-twisted action it is possible to show that the couplings $\taub_a$ as well as $\Hb$, $\Jb$ and $\Eb$ only appear in $\cQb_+$-exact terms.
One very important consequence of this for us is that in the half-twisted theory, correlators of $\cQb_+$-closed operators
are holomorphic in $J,~H$ and $E$, thus for the purpose of our computations we can set $\Jb=\Hb=\Eb=0$.

\bibliographystyle{./utphys}
\bibliography{./bigref}

\providecommand{\href}[2]{#2}\begingroup\raggedright\begin{thebibliography}{10}

\bibitem{Beasley:2003fx}
C.~Beasley and E.~Witten, ``{Residues and world-sheet instantons},'' {\em JHEP}
  {\bf 10} (2003)  065,
\href{http://arxiv.org/abs/hep-th/0304115}{{\tt arXiv:hep-th/0304115}}.

\bibitem{Dine:1986zy}
M.~Dine, N.~Seiberg, X.~G. Wen, and E.~Witten, ``{Nonperturbative Effects on
  the String World Sheet},''
{\em Nucl. Phys.} {\bf B278} (1986)  769.

\bibitem{Dixon:1987bg}
L.~J. Dixon, ``{Some world sheet properties of superstring compactifications,
  on orbifolds and otherwise},''. Lectures given at the 1987 ICTP Summer
  Workshop in High Energy Phsyics and Cosmology, Trieste, Italy, Jun 29 - Aug
  7, 1987.

\bibitem{Distler:1986wm}
J.~Distler, ``Resurrecting (2,0) compactifications,''
{\em Phys. Lett.} {\bf B188} (1987)  431--436.

\bibitem{Distler:1987ee}
J.~Distler and B.~R. Greene, ``{Aspects of (2,0) string compactifications},''
{\em Nucl. Phys.} {\bf B304} (1988)  1.

\bibitem{Berglund:1995yu}
P.~Berglund, P.~Candelas, X.~de~la Ossa, E.~Derrick, J.~Distler, {\em et al.},
  ``{On the instanton contributions to the masses and couplings of E(6)
  singlets},'' {\em Nucl.Phys.} {\bf B454} (1995)  127--163,
  \href{http://arxiv.org/abs/hep-th/9505164}{{\tt arXiv:hep-th/9505164
  [hep-th]}}.

\bibitem{Braun:2007xh}
V.~Braun, M.~Kreuzer, B.~A. Ovrut, and E.~Scheidegger, ``{Worldsheet instantons
  and torsion curves. Part A: Direct computation},'' {\em JHEP} {\bf 10} (2007)
   022,
\href{http://arxiv.org/abs/hep-th/0703182}{{\tt arXiv:hep-th/0703182}}.

\bibitem{Aspinwall:2011us}
P.~S. Aspinwall and M.~R. Plesser, ``{Elusive worldsheet instantons in
  heterotic string compactifications},''
  \href{http://arxiv.org/abs/1106.2998}{{\tt arXiv:1106.2998 [hep-th]}}.

\bibitem{Witten:1993yc}
E.~Witten, ``{Phases of N = 2 theories in two dimensions},'' {\em Nucl. Phys.}
  {\bf B403} (1993)  159--222,
\href{http://arxiv.org/abs/hep-th/9301042}{{\tt arXiv:hep-th/9301042}}.

\bibitem{Silverstein:1995re}
E.~Silverstein and E.~Witten, ``{Criteria for conformal invariance of (0,2)
  models},'' {\em Nucl. Phys.} {\bf B444} (1995)  161--190,
\href{http://arxiv.org/abs/hep-th/9503212}{{\tt arXiv:hep-th/9503212}}.

\bibitem{Basu:2003bq}
A.~Basu and S.~Sethi, ``World-sheet stability of (0,2) linear sigma models,''
  {\em Phys. Rev.} {\bf D68} (2003)  025003,
\href{http://arxiv.org/abs/hep-th/0303066}{{\tt hep-th/0303066}}.

\bibitem{Morrison:1994fr}
D.~R. Morrison and M.~Ronen~Plesser, ``{Summing the instantons: quantum
  cohomology and mirror symmetry in toric varieties},'' {\em Nucl. Phys.} {\bf
  B440} (1995)  279--354,
\href{http://arxiv.org/abs/hep-th/9412236}{{\tt arXiv:hep-th/9412236}}.

\bibitem{Distler:1995mi}
J.~Distler, ``Notes on (0,2) superconformal field theories,''
\href{http://arxiv.org/abs/hep-th/9502012}{{\tt hep-th/9502012}}.

\bibitem{Dine:1987bq}
M.~Dine, N.~Seiberg, X.~G. Wen, and E.~Witten, ``{Nonperturbative Effects on
  the String World Sheet. 2},''
{\em Nucl. Phys.} {\bf B289} (1987)  319.

\bibitem{McOrist:2008ji}
J.~McOrist and I.~V. Melnikov, ``{Summing the instantons in half-twisted linear
  sigma models},'' {\em JHEP} {\bf 02} (2009)  026,
\href{http://arxiv.org/abs/0810.0012}{{\tt arXiv:0810.0012 [hep-th]}}.

\bibitem{Aspinwall:2014ava}
P.~S. Aspinwall and B.~Gaines, ``{Rational Curves and (0,2)-Deformations},''
\href{http://arxiv.org/abs/1404.7802}{{\tt arXiv:1404.7802 [hep-th]}}.

\bibitem{Donagi:2014koa}
R.~Donagi, Z.~Lu, and I.~V. Melnikov, ``{Global aspects of (0,2) moduli space:
  toric varieties and tangent bundles},''
\href{http://arxiv.org/abs/1409.4353}{{\tt arXiv:1409.4353 [hep-th]}}.

\bibitem{Bertolini:2013xga}
M.~Bertolini, I.~V. Melnikov, and M.~R. Plesser, ``{Hybrid conformal field
  theories},'' \href{http://dx.doi.org/10.1007/JHEP05(2014)043}{{\em JHEP} {\bf
  1405} (2014)  043},
\href{http://arxiv.org/abs/1307.7063}{{\tt arXiv:1307.7063}}.

\bibitem{Bertolini:2014ela}
M.~Bertolini, I.~V. Melnikov, and M.~R. Plesser, ``{Accidents in (0,2)
  Landau-Ginzburg theories},''
\href{http://arxiv.org/abs/1405.4266}{{\tt arXiv:1405.4266 [hep-th]}}.

\bibitem{Wess:1992cp}
J.~Wess and J.~Bagger, {\em {Supersymmetry and supergravity}}.
\newblock Princeton University Press, 1992.

\bibitem{Katz:2004nn}
S.~H. Katz and E.~Sharpe, ``Notes on certain (0,2) correlation functions,''
  {\em Commun. Math. Phys.} {\bf 262} (2006)  611--644,
\href{http://arxiv.org/abs/hep-th/0406226}{{\tt hep-th/0406226}}.

\end{thebibliography}\endgroup

\end{document}